\documentclass[twocolumn,showpacs,prl,amsmath]{revtex4} 
\usepackage[dvips]{graphicx}
\usepackage{epsfig,pstricks}
\usepackage{comment}
\usepackage{amsmath}
\usepackage{amsfonts}
\usepackage{xspace}

\newcommand{\ie}{{\it i.e.}\xspace}

\newcommand{\etal}{{\it et al.}\xspace}
\newcommand{\SOC}{\text{SOC}}
\newcommand{\AS}{\text{AS}}

\newcommand{\aveX}[2]{\left\langle#1 \right\rangle_{\scriptsize #2}}
\newcommand{\aveAS}[1]{\aveX{#1}{\AS}}
\newcommand{\aveSOC}[1]{\aveX{#1}{\SOC}}

\newcommand{\varX}[2]{\sigma^{2}_{\scriptsize #2}\left(#1 \right)}

\newcommand{\stddevSOC}[1]{\varX{#1}{\SOC}}

\newcommand{\GCX}[1]{\mathcal{G}^{\scriptsize #1}}
\newcommand{\GCAS}{\GCX{\AS}}
\newcommand{\GCSOC}{\GCX{\SOC}}

\newcommand{\mfX}[1]{a^{\scriptsize #1}} 
\newcommand{\mfAS}{\mfX{\AS}} 
\newcommand{\mfSOC}{\mfX{\SOC}}

\newcommand{\elabel}[1]{\label{eq:#1}}

\newcommand{\eref}[1]{eq.~(\ref{eq:#1})}

\newcommand{\flabel}[1]{\label{fig:#1}}
\newcommand{\fref}[1]{fig.~\ref{fig:#1}}
\newcommand{\Fref}[1]{Fig.~\ref{fig:#1}}

\begin{document}

\title{Tuning- and order parameter in the SOC ensemble}

\author{O. Peters$^{1,2}$}
\email{ole@santafe.edu}
\homepage{http://www.sanatafe.edu/~ole}
\affiliation{
$^1$Dept. of Mathematics 
and Grantham Institute for Climate Change, 
Imperial College London, 
180 Queens Gate, 
London SW7 2AZ, 
UK\\
$^2$Department of Atmospheric Sciences,
University of California, Los Angeles,
405 Hilgard Avenue,
Los Angeles,
California 90095-1565,
USA}

\author{G. Pruessner}
\email{g.pruessner@imperial.ac.uk}
\homepage{http://www2.imperial.ac.uk/~pruess}
\affiliation{Department of Mathematics,
Imperial College London,
180 Queen's Gate,
London SW7 2AZ,
UK\\}

\date{\today}

\begin{abstract}The one-dimensional Oslo model is studied under
  self-organized criticality (SOC) conditions and under absorbing
  state (AS) conditions. While the activity signals the phase
  transition under AS conditions by a sudden increase, this is not the
  case under SOC conditions. The scaling parameters of the activity
  are found to be identical under SOC and AS conditions, but in SOC
  the activity lacks a pickup.
\end{abstract}


\maketitle

\section{Introduction}
Self-organized criticality (SOC) \cite{Jensen1998} is a form of
non-equilibrium critical behavior \cite{TangBak1988} where, in
contrast to traditional critical phenomena, only a separation of
timescales between external driving and internal relaxation is needed
for the system to develop into a scale invariant state. The
description of SOC as the result of feedback between an order
parameter and a tuning parameter was put forward by Tang and Bak
\cite{TangBak1988}. Later Dickman {\etal}
\cite{DickmanVespignaniZapperi1998} linked SOC to models with AS phase
transitions, where a tuning parameter, the particle density $\zeta$,
determines whether the system is in an active phase where it changes
in time or in an inactive phase where the system is stuck in one
configuration. The order parameter of these transitions is the density
of sites about to topple, called the activity. Dickman \etal~ suggested
that the critical behavior of SOC systems is numerically identical to
that of these well understood \cite{Hinrichsen2000} non-equilibrium
critical phenomena.

The intuition that the two types of critical phenomena are related can
be understood as follows: conservative SOC sandpile models can be
turned into AS systems by making the open boundaries periodic, and
ceasing to drive the systems. Thereby, for each SOC system a
corresponding AS system can be constructed. Activity in SOC systems
transports particles in a random fashion, which can lead to
dissipation at the open boundaries. The activity stops whenever the
system gets stuck in an inactive (absorbing) state. It is then
re-activated by the external drive. Thus, the SOC system is repeatedly
driven through the corresponding absorbing-state phase transition.  A
key assumption in the explanation of SOC in terms of this ``AS
mechanism'' is that the SOC activity behaves just like in AS with a
sudden pickup: ``Slow driving {\it pins} $\zeta$ at its critical
value: if it exceeds $\zeta_c$, activity is generated, and thereby
dissipation''\cite{VespignaniETAL1998}.

The main result of the present study is that this assumption is not
valid. The critical properties of the AS transition are reflected in
the finite-size scaling of the SOC activity. But as a function of the
particle density, the activity shows no sign of a phase transition
under SOC conditions.

We note that finding the critical parameters, exponents, moment ratios
etc. is incomparably easier in an SOC system than under AS
conditions. For instance, the interesting long-time limit is difficult
to access in simulations of AS transitions as the number of active
systems in an ensemble shrinks exponentially with time
\cite{deOliveiraDickman2005,Pruessner2007}.  As a numerical tool for
studying non-equilibrium phase transitions, SOC could be extremely
useful if it was fundamentally understood (as opposed to numerically
determined) which properties measured in SOC are universal.

\section{The Oslo model}
We study the Oslo model \cite{ChristensenETAL1996} in one
dimension. This model is believed to be in the C-DP universality class
(directed percolation coupled to a non-diffusive conserved field), as
is the Manna model \cite{Manna1991}. In addition to the results shown
here, all findings were confirmed for the two-dimensional Oslo model
and the one- and two-dimensional Manna models. Further details will
appear elsewhere.

\emph{Initialization in the AS version:} $\zeta L \in {\mathbb{N}}$
particles for a fixed, given density $\zeta$ are added at uniformly
randomly chosen sites in a lattice of linear size $L$ with periodic
boundary conditions (topology of a ring). Every site is assigned a
threshold particle number, either $1$ or $2$, randomly, independently
and with equal probabilities. Sites exceeding this local threshold are
called ``active'' and subject to the relaxation process described
under \emph{Toppling} below.\\
\emph{Initialization in the SOC version:} The system starts out empty
and is filled only by the \emph {External drive} (below). Every site
is assigned a random threshold as in AS.\\
\emph{Toppling:} Every active site is selected randomly with equal
probability. One particle is moved to each of the nearest neighbors,
and a new random threshold, $1$ or $2$, is assigned to the originating
site. \\
\emph{Avalanches:} The toppling is repeated until no active site is
left. Avalanches are defined as a sequence of topplings induced by a
single driving step (below).\\
\emph{Microscopic time:} Active sites topple at unit rate, which
defines the microscopic time unit.
In AS, observables are clearly time-dependent, as the
system relaxes from the initial unstable state to complete quiescence,
generating a single avalanche. The average over initial states in AS
is denoted $\aveAS{\cdot}$.  Where the time $t$ is dropped as a
parameter from the observables, time averaging with appropriate
weighting has been applied. In SOC, averaged observables are denoted
$\aveSOC{\cdot}$. This ensemble can be conditioned on the density of
particles $\zeta$.\\
\emph{Conditional activity:} The activity is the density of sites
exceeding the local particle threshold. It is the order parameter in
the sense that a non-zero asymptotic value of
$\lim_{t\to\infty}\lim_{L\to \infty}\aveAS{\rho_a(\zeta;L,t)}$
indicates the active phase. Only non-zero measurements of the
instantaneous activity contribute to estimates of the averages -- in
AS this is done by explicit conditioning (only active systems are part
of the AS ensemble), while SOC systems are re-activated immediately
whenever they fall into an absorbing state. Nonetheless, it is
important to distinguish between\\ 
1) the AS activity $\aveAS{\rho_a(\zeta;L)}$ measured at constant
(conserved) externally set $\zeta$ and\\
2) the SOC activity $\aveSOC{\rho_a(\zeta;L)}$, conditioned on the
value $\zeta$ of the fluctuating density, $\zeta(t)$, in the SOC
ensemble.\\
\emph{Boundary conditions:} In the AS version, boundaries are
periodically closed and no dissipation of particles takes place. In
the SOC version, boundaries are open, \ie toppling across a boundary
removes particles from the system.\\
\emph{External drive:} The AS models are not driven. Time series (see
below) terminate when all activity has ceased. This happens, in a
finite system, with a rate that is bounded away from zero, \ie in any
finite system the avalanche eventually comes to a halt (below a
trivial density limit). While the model in the AS version is restarted
by resetting it to a new initial condition, the SOC version is driven
externally to compensate the loss of particles at the boundary. The
driving occurs only at quiescence, thereby completely separating the
timescales of toppling and driving.\\
\emph{Time series in the AS version:} An ensemble of systems is
prepared, and observables are averaged over the members of the
ensemble (of systems that are still active) at equal times, producing
time series of fluctuating observables.
We focus here on moments of the activity in the quasi-stationary
state, $\aveAS{\rho_a^k(\zeta)}$, that is, moments of the activity
after a transient but conditional to activity
\cite{MarroDickman1999,Pruessner2007}.\\
\emph{Time series in the SOC version:} Because the SOC systems are
driven externally, they do not need to be reset to the initial state
and restarted. After a sufficient transient, which is dominated by
the time it takes the external drive to fill the system to a density
near $\zeta_c^\SOC$, the observables fluctuate about an average value.

Typical SOC observables include the avalanche size, defined as
the number of topplings, and the duration, defined as the number
of microscopic time units that elapse during an avalanche.

Studies of AS systems typically focus on stationary and dynamic
properties more commonly studied in critical phenomena, such as the
finite-size scaling of the order parameter and its fluctuations,
survival probabilities, and spreading of activity. These observables
will be referred to as AS observables.

\section{Similarities between AS and SOC}
Given the claim that corresponding observables scale identically under
SOC and AS conditions, surprisingly few direct comparisons
exist. Specifically, we are not aware of any studies of critical
scaling (in the tuning parameter) of observables under SOC
conditions. However, it has been reported that
\begin{itemize}
\item 
  the asymptotic ($L\to\infty$) particle density in an SOC system
  coincides with the critical density of the corresponding AS system
  \cite{ChessaMarinariVespignani1998,RossiPastor-SatorrasVespignani2000,DickmanETAL2000,ChristensenETAL2004}.
\item 
  the avalanche sizes measured in an AS system, that is, the response
  of a finite quiescent AS system at the critical particle density to
  the addition of a particle, obey the same finite-size scaling as
  avalanche sizes in bulk-driven SOC systems
  \cite{ChessaMarinariVespignani1998,ChristensenETAL2004}.
\item 
  scaling relations between avalanche exponents describing SOC
  observables and exponents characterising AS systems
  \cite{ChessaMarinariVespignani1998,MunozETAL1999,DickmanETAL2000,Luebeck2004}
  are apparently valid. This implies that $\aveSOC{\rho_a(L)}$ follows
  the same scaling as $\aveAS{\rho_a(\zeta_c;L)}$.
\end{itemize}
			
\section{Differences between AS and SOC}
These similarities are remarkable because the ensemble sampled by SOC
systems is very different from that sampled by AS systems.

Our main finding is illustrated in \fref{act_scaling}. Appropriate
rescaling with $L$ of $\aveAS{\rho_a(\zeta;L)}$ and $\zeta$ collapses
the curves $\aveAS{\rho_a(\zeta;L)}$. The same scaling parameters
produce a collapse of the curves $\aveSOC{\rho_a(\zeta;L)}$, but the
two collapsed curves differ. The AS activity hence follows
\begin{equation}
\aveAS{\rho_a(\zeta;L)}=\mfAS_{\rho} L^{-\beta/\nu_\perp}\GCAS(\mfAS_{\zeta} (\zeta - \zeta_c) L^{1/\nu_\perp})
\elabel{rho_a_AS_scaling}
\end{equation}
where $\mfAS_{\rho,\zeta}$ denotes metric factors
\cite{PrivmanHohenbergAharony1991}, $\GCAS$ a scaling function and
$\nu_\perp$ and $\beta$ are the usual \cite{Hinrichsen2000} critical
exponents. The SOC activity follows
\begin{equation}
\aveSOC{\rho_a(\zeta;L)}=\mfSOC_{\rho} L^{-\beta/\nu_\perp}\GCSOC(\mfSOC_{\zeta} (\zeta - \zeta_c) L^{1/\nu_\perp})
\elabel{rho_SOC_scaling}
\end{equation}
with numerically identical constants as in \eref{rho_SOC_scaling}.
Crucially, however, the scaling functions $\GCSOC$ and $\GCAS$ are
fundamentally different, with $\aveSOC{\rho_a(\zeta;L)}$ having no
powerlaw pickup. In other words, the ensemble probed in the SOC regime
lacks the {\it one} key feature that motivates the AS mechanism.

Since the scaling parameters are identical under SOC and AS
conditions, our findings do not contradict the many observations of
valid scaling relations between SOC and AS exponents. But our findings
call into question the physically appealing narrative of the AS
mechanism, where an SOC system is pushed through an AS phase
transition, into the active phase by a slow drive and strongly
repelled from there as a sudden increase in activity above $\zeta_c$
leads to dissipation. This sudden increase does not exist in SOC. Our
critique \cite{PruessnerPeters2006} was based on the disbelief that a
linear feedback loop like the equation of motion
\eref{equation_of_motion} could drive a non-linear system like the
Oslo model to a (highly non-linear) phase
transition. \Fref{act_scaling} specifies the discord between SOC and
AS ensemble, invalidating the key proposition
$\aveSOC{\rho_a}(\zeta;L)=\aveAS{\rho_a}(\zeta;L)$ underlying the AS
mechanism.

Neither the order parameter nor its variance (not shown) have special
features near the critical particle density $\zeta_c$. Whereas a
non-linear $\zeta$-dependence is clearly seen under  AS conditions,
$\aveAS{\rho_a(\zeta;L)}$, over the range of $\zeta$
sampled by the SOC system, $\aveSOC{\rho_a(\zeta;L)}$
increases approximately linearly with $\zeta$.

\begin{figure}
\includegraphics*[width=6cm,angle=-90]{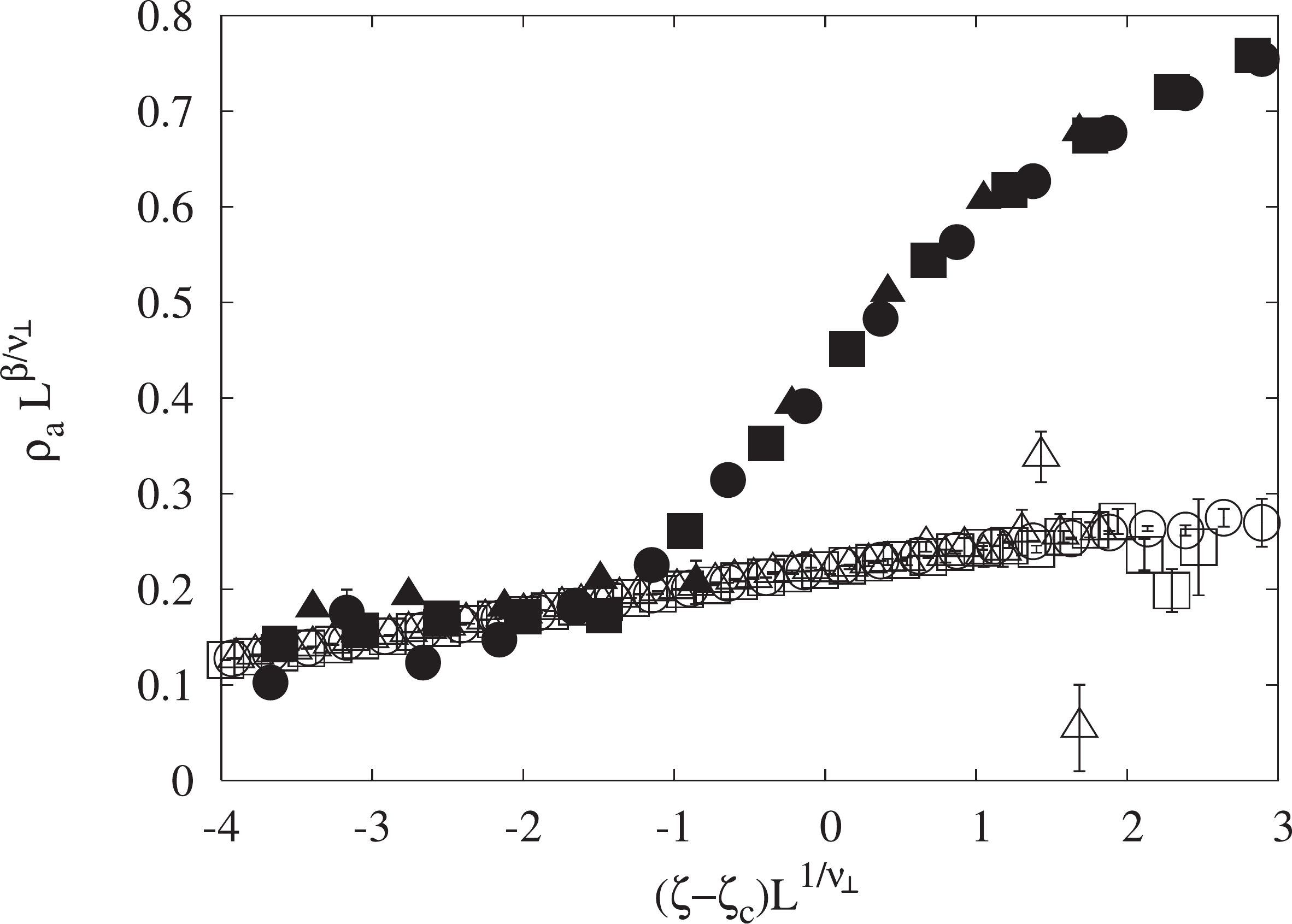}
\caption{Rescaled order parameter versus rescaled reduced density
  (circles: $L=256$, squares: $L=1024$, triangles: $L=4096$). The
  dependence on $\zeta$ of the SOC order parameter
  $\aveSOC{\rho_a}(\zeta;L)$ (open symbols), looks very different from
  the AS order parameter $\aveAS{\rho_a}(\zeta;L)$ (filled
  symbols). The same scaling parameters, $\beta=0.26, \nu_\perp=1.33$
  \cite{RamascoMunozdaSilvaSantos2004}, and $\zeta_c=1.73260$
  \cite{ChristensenETAL2004} are used for SOC systems and for AS
  systems. The active phase can be seen in AS mode as a power-law
  pick-up in the scaling function but is invisible in SOC mode. For
  the extreme values of $\zeta$ reached by the SOC system, some error
  bars are curiously small, which is due to small sample sizes at
  extreme parameter values.}  \flabel{act_scaling}
\end{figure}

The distribution of the SOC system's tuning parameter
$P(\zeta_{\SOC}(t);L)$ is well described by a Gaussian.
\Fref{zeta_width} shows its width $\stddevSOC{\zeta}$, which is a
measure for the apparent critical region in the SOC model. It is found
to shrink as $L^{-1/\nu_\perp}$, which is the minimum rate required for
universal finite size scaling in the light of the AS mechanism
\cite{PruessnerPeters2006}.

\begin{figure}
\includegraphics*[width=6cm,angle=-90]{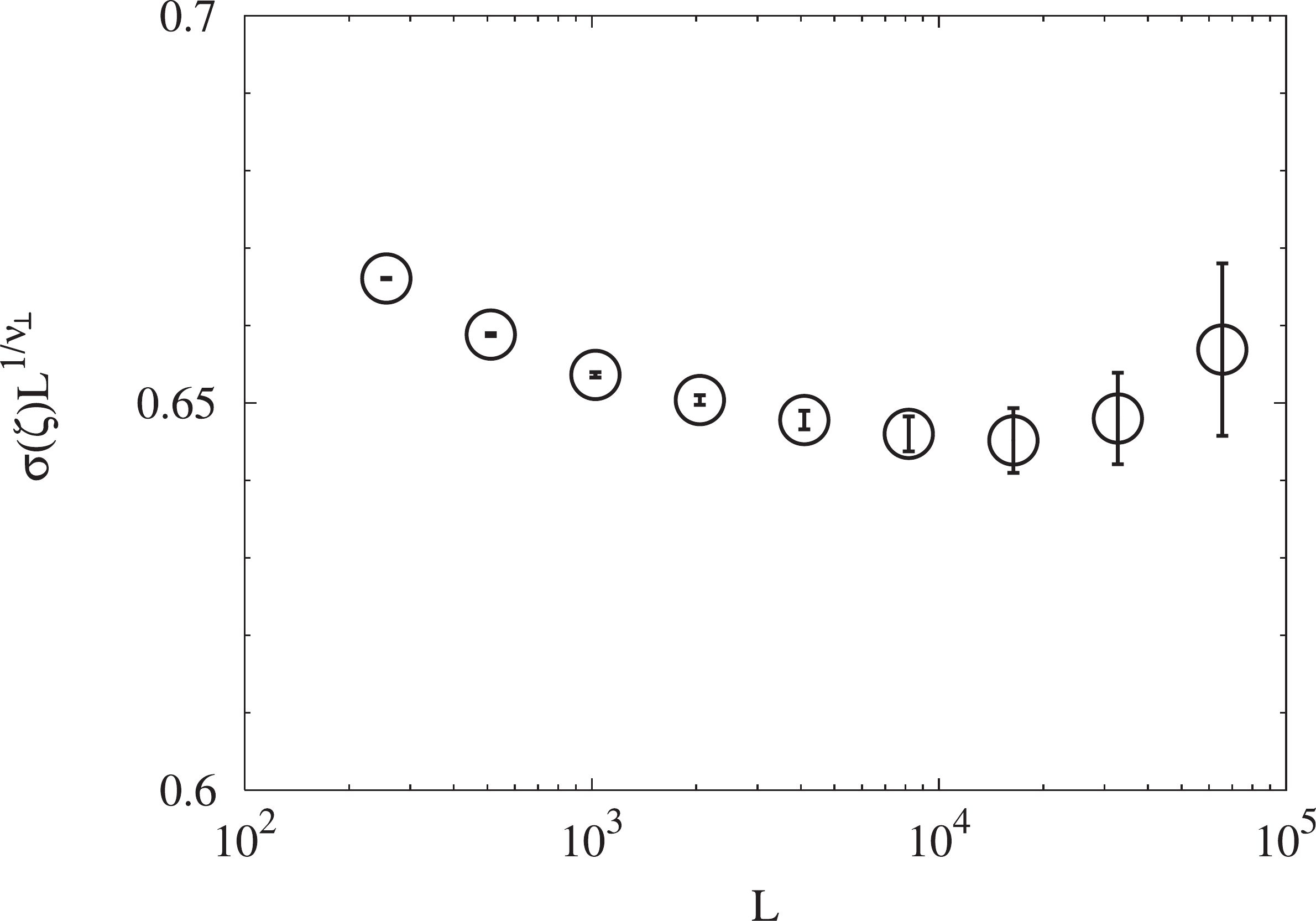}
\caption{The width of the distribution $P(\zeta_{\SOC}(t);L)$
  as a function of system size $L$ shrinks as $L^{-1/\nu_\perp}$, enabling
  the observation of universal (AS-) exponents in the SOC
  ensemble. The value $\nu_\perp=1.33$ was used for the rescaling in this
  figure. The re-scaled quantity varies by less than 4\% between
  $L=256$ and $L=65,536$.} \flabel{zeta_width}
\end{figure}

\section{Discussion}

The activity $\aveSOC{\rho_a(\zeta)}$ in the SOC model does not signal
the onset of a phase transition around $\zeta_c$ by a kink.  This
seems to contradict the widely accepted explanation of how the SOC
system finds the critical point.

The comparatively suppressed activity for $\zeta>\zeta_c$ under SOC
conditions confirms that the difference between the ensembles is
significant: In order for the SOC system to advance deep into the
high-$\zeta$ region, low dissipation and, presumably, low activity are
needed. The particle density $\zeta$ under SOC conditions thus has a
different effect from $\zeta$ under AS conditions.

Scaling laws relating SOC and AS exponents are based on the assumption
that moments $\aveSOC{\rho_a^k(L)}$ and $\aveAS{\rho_a^k(\zeta_c;L)}$
follow the same scaling. The validity of such scaling laws has been
confirmed repeatedly in the literature
\cite{LuebeckHeger2003,ChristensenETAL2004}. We have added to these
studies the direct observation of identical scaling of
$\aveSOC{\rho_a(\zeta;L)}$ and $\aveAS{\rho_a(\zeta;L)}$, \ie over a
range of $\zeta$, but we have also shown where the equivalence between
corresponding AS and SOC models breaks down.

Given the clear difference between the AS and SOC-behavior of such a
defining observable as the order parameter near criticality -- why do
we observe identical finite-size scaling and, even more puzzlingly,
why identical critical densities?  The ideal theory answering this
question includes a recipe for turning any phase transition into SOC
-- if that is only possible for some ``special'' AS transitions (maybe
there is only one such universality class), the theory has to explain
this restriction, and clarify how much of our understanding of phase
transitions applies to AS transitions at all.

The AS mechanism of self-organization, as formulated by Dickman \etal~
can be paraphrased in a mean-field equation of motion for the tuning
parameter \cite{DickmanVespignaniZapperi1998},
\begin{equation}
\partial_t \zeta(t)=h(t,L)-\epsilon(t,L) \rho_a^\SOC(t,\zeta;L),
\elabel{equation_of_motion}
\end{equation}
where $h(t;L)$ is the rate of increase in $\zeta$, measured on the
microscopic time scale, due to the addition of particles, and
$\epsilon(t;L) \rho_a^\SOC(t,\zeta;L)$ is the loss rate. Even if one
accepts that such a simple feedback loop can drive a system into a
phase transition, the models we investigated (Oslo and Manna 1-$d$ and
2-$d$) do not seem to be governed by its supposed equation of motion,
\eref{equation_of_motion}, because the detailed behavior of
$\rho_a^\SOC$ in this equation cannot be equated to that of
$\rho_a^\AS$.

Because it is so widely accepted in the literature that SOC probes a
corresponding AS phase transition
\cite{VespignaniETAL1998,ChessaMarinariVespignani1998,RossiPastor-SatorrasVespignani2000,BonachelaETAL2007},
it is often difficult to see which results are based on this
assumption and which support it independently. We emphasize that the
scaling relations between AS and SOC are not an explanation of the
equivalence between AS and SOC; their apparent validity constitutes
the observation of this equivalence that still needs to be understood.

\acknowledgments
We thank R. Dickman for sharing data from a previous study.

\bibliography{/Users/obp48/bibliography/bibliography}

\end{document}